\documentclass[a4paper,10pt]{article}
\usepackage[a4paper]{geometry}
\usepackage{amsmath}
\usepackage{amssymb}
\usepackage{xcolor}
\usepackage{comment}
\usepackage[T1]{fontenc}
\usepackage[utf8]{inputenc}
\usepackage[english]{babel}
\usepackage{cite}
\usepackage{graphicx}
\usepackage[colorlinks=true,citecolor=blue,urlcolor=blue,linkcolor=red]{hyperref} %
\usepackage{amsthm}
\newtheorem{theorem}{Theorem}
\newtheorem{conjecture}{Conjecture}

\newtheorem{proposition}[theorem]{Proposition}
\newtheorem{observation}[theorem]{Observation}
\numberwithin{equation}{section}

\definecolor{lightblue}{rgb}{0.8,0.8,1}

\title{Block-Circulant Complex Hadamard Matrices}
\author{Wojciech Bruzda\footnote{w.bruzda[at]uj.edu.pl}}
\date{{\small Jagiellonian University,\\ Faculty of Physics, Astronomy and Applied Computer Science,\\ ul. {\L}ojasiewicza 11, 30-348, Krak\'{o}w, Poland}\\
\medskip
{\small \today}}

\begin{document}
\maketitle

\begin{abstract}
A new method of obtaining a sequence of isolated complex Hadamard matrices
(CHM) for dimensions $N\geqslant 7$, based on block-circulant structures, is presented. 
We discuss, several analytic examples resulting from a modification of the Sinkhorn algorithm.
In particular, we present new isolated matrices of orders $9$, $10$ and $11$, which elements are not roots of unity, and also several
new multiparametric families of order $10$.
We note novel connections between certain eight-dimensional matrices
and provide new insights towards classification of CHM for $N\geqslant 7$.
These contributions can find real applications in Quantum Information Theory and constructions of new families of Mutually Unbiased Bases or Unitary Error Bases. 
\end{abstract}

\maketitle


\section{Complex Hadamard Matrices}

Consider a set $\mathbb{H}(N)$ of rescaled and unimodular unitary matrices $H$ of order $N>1$ such that
$HH^{\dagger}=N\mathbb{I}_N$ and $H_{jk}=\exp\{2i \pi p_{jk}\}$ for $p_{jk}\in[0,1)$
and $j,k\in\{1,2,...,N\}$. Matrix $\mathbb{I}_N$ is the $N$-dimensional identity matrix while
elements of $\mathbb{H}(N)$ are called {\sl complex Hadamard matrices} (CHM)~\cite{TZ06}.
A special subset of CHM called Butson type matrices is distinguished; ${\rm B}\mathbb{H}(N, q)\subsetneq \mathbb{H}(N)$~\cite{Bu62,Bu63}.
Each entry of a matrix $H\in{\rm B}\mathbb{H}(N, q)$ is
$q^\textrm{th}$ root of unity, that is $p_{jk} = m_{jk}/q$ for some $m_{jk}\in\mathbb{N}\cup\{0\}$, $\mathbb{N}:=\{1,2,3,...\}$.
Complex Hadamard matrices are extension of real (with only $\mp 1$ entries) Hadamard matrices~\cite{Ha93}
and can be further generalized over quaternions~\cite{CD08} or general groups~\cite{Br88,Ha97}.
Real Hadamard matrices form a special case of Butson matrices, ${\rm B}\mathbb{H}(N,2)$.
In this paper we consider mainly non-Butson type CHM.

The occurrence of complex Hadamard matrices in theoretical and experimental physics is omnipresent.
It is enough to only mention the key examples like
{\sl dense coding} and {\sl quantum teleportation schemes}~\cite{We01},
{\sl mutually unbiased bases}~\cite{DEBZ10, BBLTZ07} (MUB),
{\sl nice error bases}~\cite{Kn96}, {\sl quantum designs}~\cite{Ha97,GS81},
{\sl unitary error bases}~\cite{MV16} (UEB), {\sl unitary depolarizers}~\cite{We01, KR04}, and
{\sl tomography of quantum states}~\cite{Iv81, WF89}.
The motivation for working on CHM can be continued over many other practical and engineering applications~\cite{HKS03, Ho07},
while purely mathematical aspects of such objects are presented in the recent comprehensive monograph by Banica~\cite{Ba21}.

\section{Classification and Invariants}

It is easy to notice that $\forall\,N : \mathbb{H}(N)\neq\emptyset$ since such
an example is provided for any dimension $N$ by the Fourier matrix,
\begin{equation}
F_N=\sum_{j,k=0}^{N-1}|j\rangle\langle k|\exp\Bigg\{\frac{2 i\pi jk}{N}\Bigg\}\in{\rm B}\mathbb{H}(N,N)\subset\mathbb{H}(N).
\end{equation}
Apart from individual numerical findings, there are many systematic constructions of CHM. From simply rearranging the elements and/or expanding the size
of a given matrix~\cite{Wi44, Di04}, through
more advanced mathematical apparatus involving tiling abelian groups~\cite{MRS07}, 
graph theory~\cite{GS70, La16}, Golay sequences~\cite{LSO13}, and
orthogonal maximal abelian $*$-subalgebras~\cite{BN06}.
Further examples and references are summarized in the ``Catalog of Complex Hadamard Matrices''~\cite{CHM_catalogue}.
In this paper we will follow two different paths: a numerical method based on the Sinkhorn algorithm and
some specific constructions using particular symmetries inside the matrix.

CHM come in a variety of forms.
We distinguish three general classes: {\sl isolated points}, {\sl affine}
and {\sl nonaffine families} ({\sl orbits})~\cite{TZ06}.
Two matrices $H_1$, $H_2\in\mathbb{H}(N)$ are called {\sl equivalent}~\cite{Ha96}, written
$H_1\simeq H_2$, if there
exist two diagonal unitary matrices
$D_1$, $D_2\in\mathbb{U}(N)$ and two permutation matrices
$P_1$, $P_2$ such that
\begin{equation}
H_1 = D_1 P_1 H_2 P_2 D_2.\label{equivalence_relation}
\end{equation}
We say that $H$ is an {\sl isolated} matrix,
if its neighborhood\footnote{Matrix neighborhood
is a natural generalization of a standard point-like neighborhood
extended over $\mathbb{C}^{N\times N}$ endowed with Euclidean topology.}
contains only equivalent matrices.
Otherwise, it is a part of a family of inequivalent matrices.
The character of variability of phases $p_{jk}\in [0, 1)$ in $H_{jk} = e^{2i \pi p_{jk}}$ as
functions of orbit parameters can be linear ({\sl affine}) or nonlinear ({\sl nonaffine}).
We decorate a matrix with a superscript to indicate whether it is isolated, e.g. $S_6^{(0)}$,
or if it represents a $\delta$-dimensional orbit (orbit depending on $\delta\in\mathbb{N}$ independent parameters), like $T_8^{(1)}$.
Plethora of examples of (non)affine families can be found in~\cite{TZ06, CHM_catalogue}.

The main problem concerning CHM is their classification~\cite{TZ06, CHM_catalogue, SzPhD}. 
Having defined equivalence relation we can introduce the quotient set
\begin{equation}
\mathbb{H}(N)/_{\simeq} = [H^{(1)}]\cup [H^{(2)}]\cup ...,\label{quotient_set}
\end{equation}
so the classification of CHM boils down
to determining all equivalence classes $[H^{(j)}]$ with respect to the relation~$\simeq$.
The problem for low dimensions, $2\leqslant N \leqslant 5$, has been completely described in~\cite{Ha96,Cr91} and the first open problem arises at $N=6$.
It is widely believed that the set of six-dimensional CHM consists of a single isolated
spectral matrix~\cite{Ta04} $S_6^{(0)}$ and $4$-parameter nonaffine family~\cite{Sz12} $G_6^{(4)}$ that connects all other examples,
\begin{equation}
\mathbb{H}(6)\stackrel{?}{=}S_6^{(0)}\cup \Big\{ G_6^{(4)}\Big\}.
\end{equation}

Due to the additional degrees of freedom,
there is no simple method that can tell if two matrices are equivalent or not.
The diversity of forms for any $N\geqslant 6$ prevents us from
giving a single and simple answer to the question about the structure of $\mathbb{H}(N)$
and the task of discriminating classes of CHM is not easy indeed.
Nevertheless, there exist several methods~\cite{FG03, Ni06, Sz10} which can be used
to classify or rule out a matrix from a given subset of CHM.
We will briefly recall and use only two of them: {\sl defect of a unitary matrix} and set of {\sl Haagerup invariants}~\cite{Ha96}.

Defect of a matrix $H\in\mathbb{H}(N)$, denoted ${\bf d}(H)$,
was introduced and investigated in~\cite{TZ08, Ta18} as an algebraic tool
to solve the problem of possibility of deriving
a smooth family of inequivalent matrices from a given matrix.
Not only ${\bf d}(H)$ can serve as a binary oracle that tells whether it is possible (or not) to stem
an orbit out of $H$,
but it can also assist in identifying the equivalence classes.
Informally, we can treat ${\bf d}(H)$ as a nonnegative number associated with a matrix $H$, which provides
the information about possible independent directions on the manifold of $\sqrt{N}\mathbb{U}(N)$ that $H$
can follow preserving the property of being CHM. Non-zero value of ${\bf d}(H)$ determines the upper bound for dimension of a family that
$H$ might (but not necessarily should) belong to.
In particular, we will extensively use the following fact (Lemma 3.3 in Ref.~\cite{TZ06}):
\begin{equation}
{\bf d}(H_N)=0 \quad \Longrightarrow \quad H_N=H_N^{(0)},\label{DEFECT_implication}
\end{equation}
so, vanishing defect implies that $H_N$ cannot be part of any orbit of inequivalent matrices, and it is an isolated Hadamard matrix.
However, this is only one-way criterion and there are known examples of isolated matrices for which their defect does not vanish~\cite{MS19}.

The Haagerup set~\cite{Ha96} of $H$, denoted by $\Lambda(H)$ is defined by
all possible products of $N^4$ quartets of matrix elements
\begin{equation}
\Lambda(H) = \Big\{H_{jk}H_{lm}H_{jm}^*H_{lk}^* : j,k,l,m\in\{1,...,N\}\Big\}\label{Lambda_H},
\end{equation}
where $^*$ is complex conjugate. Both, defect and Haagerup set are invariant with
respect to equivalence relation~\eqref{equivalence_relation}, which means that
${\bf d}(H) = {\bf d}(P_1 D_1 H D_2 P_2)$ and
$\Lambda(H) = \Lambda(P_1 D_1 H D_2 P_2)$. Hence,
we can formulate another helpful criterion
\begin{equation}
\Lambda(H_1) \neq \Lambda(H_2) \quad \Longrightarrow \quad H_1 \not\simeq H_2.\label{LAMBDA_implication}
\end{equation}
Again, the converse is not true, and it is possible to
find two inequivalent matrices with perfectly coinciding sets $\Lambda$. 

Finally, we will mostly present matrices in the special {\sl dephased form} in which
first row and first column consists of ones.
Every CHM can be brought to a dephased form by multiplying by two diagonal unitary matrices.
This {\sl normalized} representation allows us to consider only permutation matrices
in the equivalence relation, and effectively work
with the $(N-1)$-dimensional {\sl core} of a given matrix.

\section{Sinkhorn Algorithm Revisited}
\label{sec:Sinkhorn_AA}

A straightforward numerical recipe to search for a new CHM is a random walk procedure, where the objective function to be optimized (minimized) reads
\begin{equation}
\mathcal{Z}(X)=||XX^{\dagger}-N\mathbb{I}_N||_{\rm F},\label{ZIEL_function}
\end{equation}
with subscript $_{\rm F}$ denoting Frobenius norm of a matrix.
For small dimensions $6\leqslant N\leqslant 16$, almost every initial (randomly chosen) matrix $X$ swiftly converges to $H$ 
such that $\mathcal{Z}(H)\approx 0$, so $H$ can be considered
as a numerical approximation of CHM with arbitrarily high precision.
Such a matrix is an input for a tedious post-processing phase of recovering its final analytical form.
This method, used in the past, led to several (re)discoveries in $\mathbb{H}(N)$ for $N\leqslant 16$.
In particular, it is very handy in the case of examination of complicated families of CHM,
where one can easily fix given pattern of phases to retrieve full functional dependencies in a nonaffine orbit~\cite{Br18}.

In this paper we recall another method which proved to be more powerful. In 1964 Sinkhorn proposed the algorithm
that was originally designed to generate random bistochastic matrices~\cite{Si64, SK67, CSBZ09}.
Suppose a matrix is characterized by at least two properties, $\mathcal{P}_j$ ($j\geqslant 1$).
They need not be complementary, nor disjoint. For example, $\mathbb{H}(N)$
can be equivalently defined as the intersection of two sets: rescaled unitary matrices and unimodular matrices;
\begin{equation}
\mathbb{H}(N) = \sqrt{N}\mathbb{U}(N) \cap \mathbb{T}(N),
\end{equation}
where $\mathbb{T}(N)$ denotes a hypertorus.
To get a matrix $X_{*}$ having all desired properties, $X_{*}\in\mathcal{P}_1\cap\mathcal{P}_2\cap...$,
we can try to perform alternating mappings (projections) $\pi_j$ onto $\mathcal{P}_j$, starting with some initial argument $X$, so that
$\pi_j(X)\in\mathcal{P}_j$.
Provided that such iterative procedure exhibits contractive behavior, it should also be convergent
to an element in a not empty subset $\mathcal{P}$, common for all $\mathcal{P}_j$,
\begin{equation}
\lim_{k\to\infty}\Big(\prod_{j=1}^k\pi_j\Big)(X)=X_{*}\in\mathcal{P}\subset\bigcap_{j}\mathcal{P}_j,
\end{equation}
where the product of $\pi_j$'s above should be understood in terms of map compositions.
In the particular case of CHM, the algorithm takes the following form:

\begin{enumerate}
\item[S1)] draw a complex matrix (a seed) $X\in\left(\mathbb{C}\setminus\{(0,0)\}\right)^{N\times N}$ at random (we exclude zeros to avoid problems in the next step),
\item[S2)] normalize (unimodularize) each entry of $X$ so $X\to X'=\pi_1(X)\in\mathbb{T}(N)$ such that $X'_{jk}=X_{jk}/|X_{jk}|$,
\item[S3)] perform polar decomposition of $X'=U\sqrt{X'^{\dagger}X'}$ to obtain (nearest) unitary matrix
$U=X'/\sqrt{X'{^\dagger}X'}$, so $X'\to X''=\pi_2(X')$ such that $X''=X'/\sqrt{X'^{\dagger}X'}$,
\item[S4)] repeat steps S2) and S3) until $\mathcal{Z}(\pi_2(\pi_1(...(\pi_1(X))...)))\approx 0$ up to
a given precision, with the objective function $\mathcal{Z}$ defined in~\eqref{ZIEL_function}.
\end{enumerate}

In short\footnote{We set the convention where $X$ is the input, while $Y$ is the output. Hence, many matrices will be called $Y_N$ for some $N$.}: $Y=$ \texttt{sinkhorn}$(X)$.
The convergence of the alternating procedure is assured in the case of convexity
of all components~\cite{BB96} $\mathcal{P}_j$.
Despite the fact that none of the subsets determining $\mathbb{H}(N)$ is convex,
the above method applied to an initial matrix with no vanishing entries
can effectively produce very accurate approximations of CHM from $\mathbb{H}(N)$ for $6\leqslant N\leqslant 16$. Hence, we
will use this algorithm as a reliable tool that demonstrated its fitness in many numerical simulations.
The advantage of this technique  is the speed and performance.
However, one seems to pay the price of loosing full control over the matrix structure, 
as it is no longer possible to easily fix particular entries of $X$ imposing a concrete appearance.
Nevertheless, as we will see, this method can produce matrices with a surprisingly high degree of internal symmetry.
Appendix~\ref{app:low_dimensions} contains several examples (including isolated cases, affine and nonaffine families)
of order $7\leqslant N\leqslant 13$ obtained in this way numerically. Many of them are new examples of complex Hadamard matrices
and (if possible) they are presented in an explicit analytical form.

In the next section we present a specific $9$-dimensional CHM obtained by the Sinkhorn method, the form
of which will trigger additional observations and further results -- a more
general construction that leads to special families of CHM.

\section{\texorpdfstring{A Novel Isolated Matrix $Y_{9C}^{(0)}$}{}}\label{novel_Y9}

As of 2022 all known CHM of order $N=9$ forming $\mathbb{H}(9)$ can be listed as follows~\cite{TZ06, MW12, BN06, Ka16}:
$F_3\otimes F_3$, $F_9^{(4)}$, $S_9^{(0)}$,
$B_9^{(0)}$, $N_9^{(0)}$ and $K_9^{(2)}$.
Also, a number of Butson type matrices~\cite{OLS20,BH_home} ${\rm B}\mathbb{H}(9,q)$ shall be recalled here to complete the picture.

We present a CHM obtained during numerical studies with the help
of the Sinkhorn algorithm described in Section~\ref{sec:Sinkhorn_AA}. Initially, all entries
were drawn at random, according to normal distribution, and such a seed $X$,
without any additional constraints, was supplied to the algorithm.
After dephasing,
the following matrix shows up\footnote{Subscript $_C$ that identifies this matrix will become clear in Appendix~\ref{app:low_dimensions}.}:
\begin{equation}
X\stackrel{\texttt{sinkhorn}}{\xrightarrow{\hspace*{1.1cm}}}Y_{9C}=\left[
\begin{array}{lllllllll}
1 & 1& 1& 1& 1& 1& 1& 1& 1\\
1 & a & d & a^{*} &  c^{*} &  b^{*} & c & b & d^{*}\\
1 & b & c & b^{*} & a & d^{*} & a^{*} & d & c^{*}\\
1 & c & b^{*} & c^{*} & d & a & d^{*} & a^{*} & b\\
1 & b^{*} & c^{*} & b & a^{*} & d & a & d^{*} & c\\
1 & d & a^{*} & d^{*} & b & c^{*} &  b^{*} & c & a\\
1 & a^{*} & d^{*} & a & c & b & c^{*} & b^{*} & d\\
1 & c^{*} & b & c & d^{*} & a^{*} & d & a & b^{*}\\
1 & d^{*} & a & d & b^{*} & c & b & c^{*} & a^{*}
\end{array}\right]\label{Y9}
\end{equation}
with:
\begin{align}
a&\approx -0.3396 + 0.9406i, \qquad b\approx -0.9635 + 0.2676i,\label{quadruplet_ab}\\
c&\approx -0.0365 + 0.9993i, \qquad d\approx +0.8396 + 0.5432i,\label{quadruplet_cd}
\end{align}
and $x^{*}=\frac{1}{x}$ denoting complex conjugate of unimodular numbers.
Rough analysis indicates that it does not belong to the Butson subset of CHM,
and the vanishing value of its defect
can be a clue that this matrix might represent a completely new class of equivalence in~\eqref{quotient_set} for $N=9$.

There are only four different constants $a$, $b$, $c$ and $d$ in $Y_{9C}$, which is a very
favorable circumstance.
To express these numbers analytically, one should solve the unitarity constraints.
They form a system of five nonlinear equations with four complex variables
\begin{equation}
\left\{
\begin{array}{r}
a+b+c+d+c.c.=-1,\\
a^2+b^2+c^2+d^2+c.c.=-1,\\
a/b+b/d+c/d+ac+c.c.=-1,\\
a/d+b/c+bc+ad+c.c.=-1,\\
a/c+ab+cd+bd+c.c.=-1,
\end{array}\label{UC_Y9}
\right.
\end{equation}
where $c.c.$ denotes complex conjugate of the preceding terms.
Possible methods of solving such systems of equations are described in Ref.~\cite{SzPhD}.
Supported by numerical software, we find that out of $128$ possible quadruplets $(a,b,c,d)$
fulfilling~\eqref{UC_Y9}, excluding those that do not meet the condition of unimodularity,
some solutions correspond to roots of unity and the rest recover the values from $Y_{9C}$.
After rearranging, reducing and simplifying~\eqref{UC_Y9}, we observe that the numbers in~\eqref{quadruplet_ab} and~\eqref{quadruplet_cd}
can be calculated as two pairs of particular roots of two monic palindromic polynomials\footnote{Monic polynomial ${\bf p}$ of degree $n$ has leading coefficient $c_n=1$. For brevity,
we write only coefficients of ${\bf p}$.
For example, the standard polynomial notation ${\bf p}(x)=1+2x+3x^2+2x^3+x^4$ corresponds to $(1,2,\underline{3},2,1)$. The middle factor is emphasised  to ease confirmation that they really form a palindromic sequence: $c_j=c_{n-j}$ for $j\in\{0,...,n\}$.} of degree $12$,
\begin{align}
{\bf p}_1(x)&=(1, -3, 9, -16, -12, 6, \underline{3}, 6, -12, -16, 9, -3, 1),\\
{\bf p}_2(x)&=(1, 6, 15, 26, 3, -24, \underline{-27}, -24, 3, 26, 15, 6, 1).
\end{align}
Remaining roots are either not unimodular or do not provide a Hadamard matrix and shall be discarded.
We exploit the well known fact from algebra of palindromic polynomials which says that
every such ${\bf p}(x)$ of even degree $2k$ can be expressed as
\begin{equation}
{\bf p}(x)=x^{k}{\bf q}\big(x+1/x\big),\label{palindromic_reduction}
\end{equation}
where ${\bf q}$ is another polynomial (in variable $y=x+1/x$) of degree $k$~(cf. Section C.2 in~\cite{SzPhD}).
Sometimes, this can shorten the form of ${\bf p}(x)$ significantly.
Indeed, in our case the problem can be reduced to 
calculating the roots of polynomials of $3^{\rm rd}$ degree
and, eventually, we arrive at\footnote{The Author is indebted to Oliver Reardon-Smith for pointing out an error in earlier calculations.}
\begin{align}
a&=\sqrt{\gamma_1^2-1}-\gamma_1,\quad b=\sqrt{\gamma_2^2-1}-\gamma_2,\\
c&=\sqrt{\gamma_3^2-1}+\gamma_3,\quad d=\sqrt{\gamma_4^2-1}+\gamma_4,
\end{align}
where
\begin{align}
\gamma_1&=\frac{1}{4}\sqrt{1-\zeta_2-\zeta_2^*}-\frac{1}{4},\qquad \gamma_2=\frac{\sqrt{2}}{2}\sqrt{1-\zeta_1-\zeta_1^*}+\frac{1}{2},\\
\gamma_4&=\frac{1}{4}\sqrt{1-\zeta_2-\zeta_2^*}+\frac{1}{4},\qquad \gamma_3=\frac{\sqrt{2}}{2}\sqrt{1-\zeta_1-\zeta_1^*}-\frac{1}{2},
\end{align}
and
\begin{align}
\zeta_1&=\frac{\omega}{2^{4/3}}\left(43-3i\sqrt{771}\right)^{1/3},\\
\zeta_2&=2^{5/3}\omega^2\left(43+3i\sqrt{771}\right)^{1/3}
\end{align}
with the phase factor
\begin{equation}
\omega=\exp\Big\{i \pi \frac{5}{3}\Big\}.
\end{equation}

Analytic expressions for $a$, $b$, $c$ and $d$ match perfectly numerical values in~\eqref{quadruplet_ab} and~\eqref{quadruplet_cd}. This formally confirms that $Y_{9C}\in\mathbb{H}(9)$ and its defect ${\bf d}(Y_{9C})=0$, so $Y_{9C}=Y_{9C}^{(0)}$ is an isolated matrix.
Obviously $Y_{9C}^{(0)}\not\in{\rm B}\mathbb{H}(9,q)$ for any $q\in\mathbb{N}$.
Since the only known isolated matrix that is not of the Butson type
is $N_9^{(0)}$ for which
\begin{equation}
\#\Lambda\left(N_9^{(0)}\right)=18<{\bf 105}=\#\Lambda\left(Y_{9C}^{(0)}\right),
\end{equation}
we can formally formally state
\begin{proposition}
Isolated matrix $Y_{9C}^{(0)}$ is a $9$-dimensional representative of the set of CHM.
\end{proposition}
The elements of the Haagerup set of $Y_{9C}^{(0)}$ can be expressed analytically, however
too complicated formulas prevent such presentation
in any compact form. Additional material can be found on Github~\cite{GITHUB}.

\section{Families of Isolated CHM}
\label{sec:CHM_LS}

In this section, we propose a systematic construction 
which might lead to infinite families of CHM with additional property of being isolated.

As we have seen, to obtain a CHM one can a priori impose certain initial configuration on entries in
a matrix, like in Eq.~\eqref{Y9},
and solve the associated system of equations representing unitarity constraints.
However, not every configuration is valid,
as there might exist some patterns for which the space of solutions yields the empty set
or the result is not ``physical'' (when numbers are not unimodular).
This technique, already presented in the previous section,
gave rise to the new matrix $Y_{9C}^{(0)}\in\mathbb{H}(9)$. We will show that
particularly chosen configuration of entries
provides several other previously unknown CHM of orders $N\geqslant 9$,
and additionally may set the ground for more general conjectures.

\subsection{\texorpdfstring{Family of Complex Hadamard Matrices $L_N$}{}}

Consider again the very first form of $Y_{9C}^{(0)}$~\eqref{Y9}.
After permuting its rows and columns we can write it equivalently as
\begin{equation}
Y_{9C}^{(0)}\simeq Y_{9C}^{'(0)} =
\left[
\begin{array}{l|ll|ll|ll|ll}
1 & 1 &1 &1 &1 &1 &1 &1 &1\\
\hline
1 & a & a^{*}&  b & b^{*} & c & c^{*} & d & d^{*}\\
1 & a^{*} & a & b^{*} & b & c^{*} & c & d^{*} & d\\
\hline
1 & b & b^{*} & d & d^{*} & a^{*} & a & c & c^{*}\\
1 & b^{*} & b & d^{*} & d & a & a^{*} & c^{*} & c\\
\hline
1 & c & c^{*} & a^{*} & a & d^{*} & d & b^{*} & b\\
1 & c^{*} & c & a & a^{*} & d & d^{*} & b & b^{*}\\
\hline
1 & d & d^{*} & c & c^{*} & b^{*} & b & a^{*} & a\\
1 & d^{*} & d & c^{*} & c & b & b^{*} & a & a^{*}
\end{array}\right]\label{Y9_LS_form},
\end{equation}
where grid lines additionally accent the internal structure.
Suddenly, the core of the matrix resembles a coarse-grained symmetric Latin square with additional complex conjugates ($^*$) in few places,
\begin{equation}
\text{core}\Big(Y_{9C}^{'(0)}\Big)=\left[
\begin{array}{llll}
\fbox{\texttt{A}} \ \, & \fbox{\texttt{B}} & \fbox{\texttt{C}} & \fbox{\texttt{D}}\\
\\
\fbox{\texttt{B}} & \fbox{\texttt{D}} & \fbox{\texttt{A}}^* & \fbox{\texttt{C}}\\
\\
\fbox{\texttt{C}} & \fbox{\texttt{A}}^* & \fbox{\texttt{D}}^* & \fbox{\texttt{B}}^*\\
\\
\fbox{\texttt{D}} & \fbox{\texttt{C}} & \fbox{\texttt{B}}^* & \fbox{\texttt{A}}^*
\end{array}\right],
\end{equation}
where every $\fbox{\texttt{X}}$ corresponds to a $2\times 2$ matrix
\begin{equation}
\fbox{\texttt{X}}=\left[\begin{array}{ll}
x & x^*\\
x^* & x
\end{array}\right]\label{2x2_X}
\end{equation}
for $(x,\texttt{X})\in\{(a,\texttt{A}),(b,\texttt{B}),(c,\texttt{C}),(d,\texttt{D})\}$.
Looking at this structure, let us propose a formal construction
and then test it for several dimensions.

Define
\begin{equation}
L_N=\left[\begin{array}{ll}
1 & 1^{1\times (N-1)}\\
\\
1^{(N-1)\times 1} & \text{core}(L_N)
\end{array}\right]\in\mathbb{C}^{N\times N},\label{LS_pattern}
\end{equation}
where $1^{a\times b}$ denotes vectors of ones of size $a\times b$
and
\begin{equation}
\text{core}(L_N)=\text{circ}\left(
\left[\begin{array}{ll}
c_0 & c_0^*\\
c_0^* & c_0
\end{array}\right],
\left[\begin{array}{ll}
c_1 & c_1^*\\
c_1^* & c_1
\end{array}\right],
...,
\left[\begin{array}{ll}
c_{n-1} & c_{n-1}^*\\
c_{n-1}^* & c_{n-1}
\end{array}\right]
\right) \ : \ c_j\in\mathbb{C}
\label{coreLN}
\end{equation}
forms a submatrix being
a $(2\times 2)$-block circulant array of size $N-1$ with $n=(N-3)/2$. By construction $N$ must be odd, because the core always consists of even number of rows and columns.
Note that we do not introduce complex conjugates
in blocks, as it was in the case of $Y_{9C}^{'(0)}$~\eqref{Y9_LS_form}.
The question is: For what values of $N$, the matrix $L_N$ can be a member of the set $\mathbb{H}(N)$ of complex Hadamard matrices?

Given $N$, we can immediately build $L_N$, write unitarity constraints and try to solve them.
They can be rearranged to form the following system:
\begin{equation}
\left\{
\begin{array}{r}
\displaystyle{\sum_{j=0}^{2k}}\cos\alpha_j=-\frac{1}{2},\\
\displaystyle{\sum_{j=0}^{2k}}\cos \left(2 \alpha_j\right)=-\frac{1}{2},\\
k \ \text{equations with $0\leqslant i<k$} \quad : \quad \displaystyle{\sum_{j=0}^{2k}}\cos \left(\alpha_j+\alpha_{(j + 1 + i) \bmod (2k+1)}\right)=-\frac{1}{2},\\
k \ \text{equations with $0\leqslant i<k$} \quad : \quad\displaystyle{\sum_{j=0}^{2k}}\cos \left(\alpha_j-\alpha_{(j + 1 + i) \bmod (2k+1)}\right)=-\frac{1}{2},\\
\end{array}\label{TRIGONOMETRIC_LN}
\right.
\end{equation}
where $\alpha_j\in\mathbb{R}$ are phases associated with appropriate blocks in Eq.~\eqref{coreLN}.
It turns out that this particular construction stemming from the block-circulant arrays
provides CHM only under one additional restriction.
Namely, we can find a solution only in every second odd dimension, $N=3+4k$ for $k\in\mathbb{N}\cup\{0\}$.
Due to circulant symmetries, the total number of possible equations forming a
nonlinear system associated with unitarity constraints can be reduced to only $2(k+1)=(N+1)/2$ trigonometric equations with $2k+1=(N-1)/2$ real variables; $k \geqslant 0$.
This significant simplification allows us to solve such systems numerically for $N$ ranging from $11$ to $127$ with a very high precision,
and initial analysis shows that in each case the defect of $L_N$ is zero and $L_N\not\in{\rm B}\mathbb{H}(N,q)$ for any $1<q\leqslant 2^{16}$.

Obviously for $k=0$ we recover the well known Fourier matrix $F_3^{(0)}$ as well as $F_7^{(0)}$ for $k=1$. But for $k\geqslant 2$, which implies $N\geqslant 11$, all constructed matrices are believed to be new as the
fact of being isolated non-Butson matrices automatically excludes them from many known families.
Additionally, computer assisted proof confirms that the cardinalities of the Haagerup invariants~\eqref{Lambda_H} are unique.
For $N=11$ we have:
\begin{align}
\#\Lambda\left(C_{11A}^{(0)}\right) = \#\Lambda\left(C_{11B}^{(0)}\right) &= 5,\\
\#\Lambda\left(N_{11A}^{(0)}\right) = \#\Lambda\left(N_{11B}^{(0)}\right) = \#\Lambda\left(N_{11C}^{(0)}\right) &= 10,\\
\#\Lambda\left(F_{11}^{(0)}\right) &= 11,\\
\#\Lambda\left(Q_{11A}^{(0)}\right) = \#\Lambda\left(Q_{11B}^{(0)}\right) &= 63,\\
\#\Lambda\left(L_{11}^{(0)}\right) &= \bf 191,
\end{align}
and similarly for $N=15$
\begin{align}
\#\Lambda\left(A_{15A}^{(0)}\right) = \#\Lambda\left(A_{15B}^{(0)}\right) &= 5,\\
\#\Lambda\left(A_{15G}^{(0)}\right) = \#\Lambda\left(A_{15H}^{(0)}\right) &= 38,\\
\#\Lambda\left(L_{15}^{(0)}\right) & = \bf 463,
\end{align}
where $C_{11A,B}$, $N_{11A,B,C}$, $Q_{11A,B}$ and $A_{11A,B,G,H}$ represent several well known classes of complex Hadamard matrices~\cite{CHM_catalogue}.

There is no need to check invariants in higher dimensions because,
to the best of our knowledge, no isolated matrices
of this size (apart of Butson matrices~\cite{MW12, OLS20, BH_home}) were ever presented in the literature.
All these observations allow us to formulate
\begin{conjecture}
For any $k\in\mathbb{N}\cup\{0\}$ and $N=3+4k$, there exists an isolated complex Hadamard matrix
$L_N^{(0)}\in\mathbb{H}(N)$, which is not of the Butson type.\label{LN_conjecture}
\end{conjecture}

Before we comment possible implications of this conjecture,
in the next section we shall introduce even simpler method producing a sequence of matrices with very similar properties.

\subsection{\texorpdfstring{Family of Complex Hadamard Matrices $V_N$}{}}\label{CHM_VN}

Instead of struggling with a circulant core, let us simply analyze entire circulant matrix
\begin{equation}
V_N=\text{circ}\left[c_0,c_1,c_2,...,c_{N-1}\right]\in\mathbb{C}^{N\times N} \quad : \quad c_j\in\mathbb{C}.
\end{equation}
Unusual undephased representation of $V_N$ provides a compact form
of the unitarity constraints:
\begin{equation}
\sum_{j=0}^{N-1}\frac{c_j}{c_{(j+k)\bmod N}}=0 \qquad \text{for} \qquad k\in\{1,2,...,N-1\}\label{UC_LS_undephased}
\end{equation}
which is the equivalent representation of the well known problem of {\sl cyclic $N$-roots}~\cite{BF91}.
Equation~\eqref{UC_LS_undephased} was investigated by Gabidulin and Shorin in the context of autocorrelation functions~\cite{GS02}, however the authors did not relate it directly to the field of CHM. The problem of circulant matrices was also taken in details in the doctoral thesis of Sz\"oll\H{o}si~\cite{SzPhD}.
Yet another variation about circulant matrices, namely real Hadamard matrices with two circulant cores is presented in Ref.~\cite{KKS06}.

We checked that the above $N-1$ nonlinear equations for $c_j\in\mathbb{C}$ can be solved
in any dimension $6\leqslant N\leqslant 64$, which
results in a sequence of isolated and non-Butson matrices.
We also claim that this can be solved in any dimension $N\in\{6,7,8,...,64,...\}$ yielding such objects.
As we mentioned at the very beginning, all solutions reproducing Butson type matrices are intentionally discarded.
Instead we focus on checking how
far we can go with $N$ still getting isolated matrices, to draw
\begin{conjecture}
For any $N\geqslant 6$ the solution of Eq.~\eqref{UC_LS_undephased} contains at least one isolated CHM matrix $V_N^{(0)}$, which
is not of the Butson type.\label{VN_conjecture}
\end{conjecture}
While for the previous sequence of $L_N$ numerical simulations
suggest that each (proper) solution leads only to an isolated matrix,
for $V_N$ the defect is not strictly restricted to zero. We observed such a situation already for $N=8$, and
also for some slightly higher orders. However, the bigger $N$, the less chance of hitting any potential
representative of a multidimensional family. The case of $N=8$ will be explained in details in Appendix~\ref{app:low_dimensions}.

\section{Summary}

In this paper we extended the set of complex Hadamard matrices, which
can be used for diverse purposes in constructing novel and crucial objects relevant for quantum theory. Using the Sinkhorn algorithm adapted to the realm of CHM,
we introduced several isolated points including $Y_{9A,B,C}^{(0)}$, $Y_{10A}^{(0)}$, $L_{11}^{(0)}$,
and families, e.g. $T_{8A}^{(1)}$, $T_{8B,C}^{(3)}$, $Y_{10C}^{(1)}$,
$Y_{10D}^{(2)}$, $Y_{10E}^{(3)}$ and $Y_{10F}^{(4)}$ -- see Appendix~\ref{app:low_dimensions}. We also mentioned numerical examples calculated with high precision, which may serve as potential candidates for further examination.

Based on the numerical evidence we put forward two conjectures concerning existence
of a sequence of isolated CHM in infinitely many dimensions.
In each conjecture we consider only two possible patterns of CHM.
Certainly, this does not exhaust the whole range of other possibilities, in particular those, which
can be enriched with additional internal symmetries. 
For a moment we cannot see how the immensity of possible configurations
could be classified into any systematic or coherent way.

Classification of complex Hadamard matrices in large dimensions must be conducted methodically.
As the problem rapidly complicates for $N\geqslant 16$, it is necessary to impose some regularities
in order to reduce the available space on the manifold of (rescaled) unitary matrices.
Presented approach allows one to simply examine the following procedure:
given a structure with predefined symmetries,
one can solve the associated system of unitarity constraints to check  whether this provides a CHM.
Equivalently, one can use the Sinkhorn algorithm to obtain (generally) the same result numerically.
In Appendix~\ref{app:low_dimensions} we discuss only a narrow excerpt of many possible ways.

The particular curiosity of this work is that the vast majority of resulting matrices are isolated.
There are roughly three different methods of construction isolated CHM known to us.
The first one is based purely on numerical approach.
Second one follows from the theorem, which says that for any prime dimension $p$ the Fourier matrix is
isolated~\cite{TZ06};
\begin{equation}
N=p \quad \Longrightarrow \quad F_p=F_p^{(0)}.\label{Fp_isolated}
\end{equation}
The third one has its source in the construction described by McNulty and Weigert in Theorem 3 in Ref.~\cite{MW12}.
If the postulated conjectures were proven, a new way to get an infinite sequence of isolated CHM would be found.
But even if they fail to be true, we are left with two manageable constructions that so far generated dozens of genuinely
new examples of CHM in a number of dimensions $N\geqslant 8$.

One possible physical motivation for searching isolated matrices  is the straightforward connection of CHM with mutually unbiased bases (MUB).
It is much easier to examine multidimensional families and check if their representatives form possible sets of MUB.
However, it is not known whether the complete set of MUB in $\mathbb{C}^6$ or $\mathbb{C}^{10}$ can be built of (yet unknown)
isolated\footnote{Actually, it is unlikely
that for $N=6$ new isolated matrices will be discovered.}
CHM.
Having a straightforward method of obtaining many new such objects one can significantly widen the possible area of research.
Presented facts may also find applications in several branches of modern physics, including creation of
unitary error bases~\cite{MV16} (UEB)
or being used in entanglement detection~\cite{SHBAH12,HL14}.
As it has been recently shown, even incomplete (unextendible) set of MUB can be more effective to detect entanglement~\cite{HMBRBC21}. 

\section{Acknowledgments}
Author thanks Veit Elser, Markus Grassl, Dardo Goyeneche, Wojciech Tadej and Karol \.{Z}ycz\-kow\-ski for many inspiring discussions.
This work is partially funded by Foundation for Polish Science 
under the Team-Net project No.\ POIR.04.04.00-00-17C1/18-00.

\appendix

\section{\texorpdfstring{Comments on $\mathbb{H}(N)$ for some initial dimensions}{}}
\label{app:low_dimensions}

In the appendix we include several observations
concerning the structure of the set of complex Hadamard matrices in a few initial dimensions,
where it was possible to find something allegedly new, which additionally allowed us to make a concise presentation.
Due to algebraic complications, we restrict in some cases only to numerical considerations or implicit solutions.
If not stated differently, one should assume that the method used to obtain a given matrix was the Sinkhorn algorithm.
In every case in the following sections the seed set as a starting point was a random Gaussian matrix,
without imposing any additional internal symmetries. As we will observe, even in dimensions as small as $8\leqslant N\leqslant 11$,
still there are many gaps and unknowns.
Concerned Reader should check the Github repository~\cite{GITHUB} for scripts (Matlab and Mathematica)
generating all the matrices described below, also including those which are not necessarily new.

In Tables:~\ref{tab:CHM_8},~\ref{tab:CHM_9},~\ref{tab:CHM_10} and~\ref{tab:CHM_11_16} we shortly summarize particular dimensions exposing the most interesting objects. They certainly do not saturate all possibilities and do not even pretend to be any attempt to a full classification. 
Even the selected matrices are not going to be examined in details. For each matrix being a result of the Sinkhorn algorithm or a solution of $L_N$ or $V_N$ family, we provide
its generic values of defect, cardinality of Haagerup set, and possible or actual relations with
other matrices. 

\subsection{\texorpdfstring{Set $\mathbb{H}(7)$}{}}

For $N=7$ we can recover all known\cite{CHM_catalogue} $7$-dimensional CHM: $F_7^{(0)}$, $P_7^{(1)}$, $C_{7A,B,C,D}^{(0)}$ and $Q_7^{(0)}$.
In particular, during examination of numerical data, we readily discovered what we believe is an alternative analytical approach to the circulant matrix $C_{7D}$ found previously by Haagerup~\cite{Ha96}.

Let $x_{\star}$ be the maximal real root of the polynomial $1+12x+18x^2-64x^3-96x^4+45x^5+43x^6=0$.
Define
\begin{equation}
a=\frac{-1}{2x_{\star}}+i\left(\frac{1}{2}\sqrt{2\sqrt{21}+6}\cos\frac{3\pi}{14}-\sqrt{2\sqrt{21}-6}\sin\frac{3\pi}{14}-\frac{3}{4}\sqrt{2\sqrt{21}-6}\right)
\end{equation}
and consider a special pair $(b,c)$ of unimodular solutions of the following system
\begin{equation}
\left\{\begin{aligned}
1 + a + b + c + \frac{1}{a} + \frac{1}{b} + \frac{1}{c}  & = 0, \\
1 + c + \frac{1}{c} + ab + \frac{1}{ab}+bc  + \frac{1}{bc}  &= 0, \\
\end{aligned}\right.
\end{equation}
which approximate values are $b=\exp\{1.3562i\}$ and $c=\exp\{1.9006i\}$ -- compare with Eq.(112) and Eq.(113) in Ref.~\cite{TZ06}.
Writing auxiliary variables 
\begin{align}
    p&=\frac{-a}{a+b+ab(1+a+b+c)},\\
    q&=\frac{1}{1+a}\left(1+a+\frac{a^2}{c}-\frac{a}{bc}+\frac{1}{b^2c}-\frac{a^2}{c^2}-\frac{1}{bc^2}+\frac{a}{b^2c^2}\right),\\
    r&=\frac{-1}{1+a}\left(1+a+\frac{1}{c}+\frac{1}{bc}+\frac{1}{b^2c}\right),
\end{align}
the symmetric core of the isolated matrix $C_{7D}^{'(0)}$ becomes
\begin{equation}
{\rm core}\left(C_{7D}^{'(0)}\right)=\left[\begin{array}{cccccc}
 a^2 & a b & a c & a & a c^* & ab^*\\
 ab & bc & c & c^* & (bc)^* & (ab)^*\\
 ac & c & b^* & acr & cr & (b^{2})^{*}\\
 a & c^* & acr & r & ar & (bc)^*\\
 ac^* & (bc)^* & cr & ar & cq & b^*\\
 ab^* & (ab)^* & (b^{2})^{*} & (bc)^* & b^* & p
\end{array}
\right]={\rm core}\left(C_{7D}^{'(0)\rm T}\right).
\end{equation}
Changing variables from $a$, $b$, $c$ to $A=a$, $B=ab$ and $C=abc$ we can express $C_{7D}^{'(0)}$ in a generic and fully circulant form
as it originally emerged as an output from the Sinkhorn routine;
$C_{7D}^{'(0)}={\rm circ}\left([1,A,B,C,C,B,A]\right)$.
Matrix $C_{7D}^{'(0)}$ is probably the well known example of $C_{7D}^{(0)} \simeq C_{7D}^{'(0)}$.


\subsection{\texorpdfstring{Set $\mathbb{H}(8)$}{}}

Veit Elser found the matrix $V_8^{(0)}$
quite accidentally as a by-product of
computations with a completely different purpose~\cite{VE_private}.
It has been identified as an isolated point, put into the Catalog of CHM~\cite{CHM_catalogue} and apparently forgotten for years.
Its nice undephased structure reads
\begin{equation}
V_8^{(0)}=\left[
\begin{array}{rrrrrrrr}
-1 & -1 &  b &  b &  c &  c &  a &  a \\
-1 &  b & -1 &  c &  b &  a &  c & -a \\
 b & -1 &  c & -1 &  a &  b & -a &  c \\
 b &  c & -1 &  a & -1 & -a &  b & -c \\
 c &  b &  a & -1 & -a & -1 & -c &  b \\
 c &  a &  b & -a & -1 & -c & -1 & -b \\
 a &  c & -a &  b & -c & -1 & -b & -1 \\
 a & -a &  c & -c &  b & -b & -1 &  1 \\
\end{array}\right].
\end{equation}
The shape of the characteristic bands of symbols/numbers revolving symmetrically around the diagonals
reflects the obscure properties of the Sinkhorn algorithm that was used to find this matrix.
The only difference in the Elser's version of the routine was that,
to get a projection onto the set of unitary matrices of order eight, $\mathbb{U}(8)$,
QR decomposition was used instead of the polar one.

Let us quickly present its analytic form.
Unitarity constraints for $V_8^{(0)}$,
\begin{equation}
\left\{\begin{aligned}
-\frac{1}{b} - b + \frac{b}{c} + \frac{c}{b} + \frac{c}{a} + \frac{a}{c} & = 0, \\
-\frac{1}{b} -b - \frac{1}{c} -c   - \frac{c}{a} - \frac{a}{c}+ \frac{b}{a} + \frac{a}{b} & = 0, \\
 \frac{1}{c} + c + \frac{1}{a} + a+ \frac{b}{a} + \frac{a}{b} & = 0,
\end{aligned}\right.
\end{equation}
can be rearranged as in the case of $Y_{9C}^{(0)}$~\eqref{Y9} and presented in the form of a monic palindromic polynomial ${\bf p}$,
so the set of roots of ${\bf p}$ contains parameters $\{a, b,c\}$.
Polynomial ${\bf p}$ appears to be
a bit complicated
and its degree is as much as $16$. Thus, to save space,
we show only the first half of its $17$ coefficients (using the polynomial notation introduced in Section~\ref{novel_Y9})
\begin{equation}
{\bf p}(x)=(1,16,64,16,-332,-1040,-1984,-2832,\underline{-3194},...).
\end{equation}
Fortunately, it can be reduced to an octic polynomial
\begin{equation}
{\bf q}(x)=(-16,256,-96,-896,-696,-96,56,16,1),
\end{equation}
which can be further factorized into the product
\begin{align}
{\bf q}(x)=&\left(x^4+8x^3+\alpha_0x^2+\beta_0 x+\gamma_0\right)\times\nonumber\\
\times&
\left(x^4+8x^3+\alpha_1x^2+\beta_1 x+\gamma_1\right),
\end{align}
where
\begin{align}
\alpha_{\mu}&=-4+2(-1)^{\mu}\sqrt{116-2\sqrt{2}},\\
\beta_{\mu}&=-16+8(-1)^{\mu}\sqrt{10-\sqrt{2}},\\
\gamma_{\mu}&=-4\sqrt{2}+4-4(-1)^{\mu}\sqrt{4-2\sqrt{2}},
\end{align}
and $\mu\in\{0,1\}$.
Applying formulas for roots of a quatric polynomial and taking
into account all changes of variables,
we can obtain analytically all the triplets
that fully determine $V_8^{(0)}$.
We conclude with the following
\begin{observation}
$V_8^{(0)}$ is an isolated point such that
$V_8^{(0)}\not\in{\rm B}\mathbb{H}(8,q)$ for any $q\in\mathbb{N}$.
\end{observation}

Moreover, the sequence $V_N$ presented\footnote{Coincidence of the family name $V_N$ and the matrix $V_8^{(0)}$ is intentional.} in Section~\ref{CHM_VN}
tightly relates $V_8^{(0)}$ with other $8$-dimensional matrices.
The unitarity constraints~\eqref{UC_LS_undephased} for $N=8$
include the solution for matrices equivalent to
$V_8^{(0)}$, $A_8^{(0)}$, possibly~\cite{MRS07} $S_8^{(4)}$ and perhaps some other elements of $\mathbb{H}(8)$.
Before $A_8^{(0)}$ was recognised as the solution of the system~\eqref{UC_LS_undephased},
it was found numerically
using a random walk procedure over
symmetric matrices~\cite{Br18}, $A_8^{(0)}=A_8^{(0)\rm T}$.

\begin{table}[ht!]
\centering
\begin{tabular}{||l|c|l|l|l|c||}
 matrix &  ${\bf d}$ &  $\#\Lambda $ & source & comment & status\\
\hline
$Y_{8A}^{(0)}$                      & 0 & 10  & $\texttt{SV}$ & $\simeq A_8^{(0)}$ & known\\
$Y_{8B}^{(0)}$                      & 0 & 70  & $\texttt{SV}$ & $\simeq V_8^{(0)}$ & known\\
\hline
$T_{8A}^{(1)}(p_1)$                 & 3 & 10  & $\texttt{S}$  & $\stackrel{?}{\subset} T_8^{(1)}$, $T_{8A}^{(1)}=T_{8A}^{(1){\rm T}}$ & unknown\\
$T_{8B}^{(3)}(p_1,p_2,p_3)$         & 3 & 74  & $\texttt{S}$  & $\stackrel{?}{\supset} T_8^{(1)}$ & new\\
$T_{8C}^{(3)}(p_1,p_2,p_3)$         & 3 & 130 & $\texttt{S}$  & $\stackrel{?}{\supset} T_8^{(1)}$ & new\\
\hline
$G_{8A}^{(3)}(p_1,p_2,p_3)$         & 5 & 42  & $\texttt{S}$  & $\subset F_8^{(5)}$ & known\\
$G_{8B}^{(5)}(p_1,...,p_5)$ & 5 & 74  & $\texttt{S}$  & $\simeq F_8^{(5)}$  & known\\
$G_{8C}^{(5)}(p_1,...,p_5)$ & 5 & 82  & $\texttt{S}$  & $\simeq F_8^{(5)}$  & known\\
\hline
... & ... & ... & ... & ... & ...\\
\end{tabular}
\caption{Selected outputs for $N=8$ obtained from the Sinkhorn algorithm (\texttt{S}) and $V_N$ family (\texttt{V}) defined in Section~\ref{CHM_VN}.}
\label{tab:CHM_8}
\end{table}

Another matrix called $T_{8B}^{(3)}$ has the form
\begin{equation}
T_{8B}^{(3)}(p_1,p_2,p_3) = \left[\begin{array}{rrrrrrrr}
        1  &      1   &   1   &    1   &    1    &   1   &  1   &    1    \\
        1  &     -1   &   c   &   -c   &    d    &  -d   &  cd  &   -cd   \\
        1  &      a   &  -a   &   -1   &    ad   &   d   & -d   &   -ad   \\
        1  &     -a   &  -ac  &    c   &    a    &  -1   & -c   &    ac   \\
        1  &      b   &  -c   &   -bc  &   -1    &  -b   &  c   &    bc   \\
        1  &     -b   &  -1   &    b   &   -d    &   bd  &  d   &   -bd   \\
        1  &      ab  &   ac  &    bc  &   -ad   &  -bd  & -cd  &   -abcd \\
        1  &     -ab  &   a   &   -b   &   -a    &   b   & -1   &    ab   \\
    \end{array}\right],
\end{equation}
where
$a=\exp\{2i\pi p_1\}$, $b=\exp\{2i\pi p_2\}$, $c=\exp\{2i\pi p_3\}$ and
$d = \frac{1 + ab + bc + ca}{a + b + c + abc}$.
Provided that phases $p_j\in[0,1)$ are chosen so that $d$ is a valid unimodular number,
matrix $T_{8B}^{(3)}\in\mathbb{H}(8)$. Generic value of defect might suggest
the intersection of $T_{8B}^{(3)}$ with $T_{8}^{(1)}$ --
a one-parameter nonaffine family described in Ref.~\cite{Br18}, where
we anticipated that this might be a part of a bigger three-parameter structure, what is potentially
allowed by the value of its defect. 

\medskip

Next matrix reads
\begin{equation}
T_{8C}^{(3)}(p_1,p_2,p_3) = \left[\begin{array}{rrrrrrrr}
                1 & 1     & 1     &  1    & 1      &  1     &  1     &  1    \\
                1 & c_1   & c_3   & -c_3  & e      & -1     & -c_1   & -e    \\
                1 & a     & c_2   &  c_3  & b      &  a/c_1 &  c     &  d    \\
                1 & a/b   & a     &  a/d  & a/c_2  &  a/c_3 &  c_1   &  a/c  \\
                1 & a/c_2 & b     &  f    & eg     &  g     &  eg/f  &  e    \\
                1 & a/d   & c     & -a/d  & g      & -g     & -c     & -1    \\
                1 & a/c   & a/c_1 & -f    & f      & -a/c_1 & -1     & -a/c  \\
                1 & a/c_3 & d     & -1    & eg/f   & -a/c_3 & -eg/f  & -d    \\
\end{array}\right],
\end{equation}
with $a$, $b$ and $c$ as above, and
\begin{align}
            d &= -(c_3 + a + c + b + 1 + c_2 + a/c_1),\\
            e &=  (c_3/c_2 - c_1/c) / (1/d - 1/b),\\
            f &=  (a/c_2 - a/c) / (c_1/c_3 - d/b),\\
            g &= -(1 + a/c_2 + b + f + e) / (e/f + e + 1)
\end{align}
where $c_1$, $c_2$ and $c_3$ are three additional unimodular variables.
Irrespective of how $c_j$ are fixed, one can always find numerically 
such a triplet $(a,b,c)=(a(c_j), b(c_j), c(c_j))$ that $T_{8C}^{(3)}\in\mathbb{H}(8)$ with generic defect ${\bf d}=3$
and $\#\Lambda=130$.
There is yet another variant of $T_{8C}^{(3)}$, not included here, with the same  characteristics, $({\bf d},\#\Lambda)=(3,130)$, however
with much more complicated internal dependencies between entries.
One can also numerically recover a sequence of matrices with cardinalities of
their $\Lambda$-sets being close (or equal) to $170$, $178$, $242$, and several bigger values -- all of them having ${\bf d}=3$.
These matrices can be generated by the Sinkhorn algorithm alas
cannot be presented in any simple form, so they are left
for a possible future investigation. All the matrices $T_{8...}^{(3)}$ support
the claim that $T_{8}^{(1)}$ might intersect or even be included in some $T_{8...}^{(3)}$, which is yet to be fully recognized.


\subsection{\texorpdfstring{Set $\mathbb{H}(9)$}{}}

Numerical data suggests that we are also far from a complete description of $\mathbb{H}(9)$.

A symmetric matrix $Y_{9A}^{(0)}$ can be implicitly presented in the following form.
\begin{align}
&{\rm core}\left(Y_{9A}^{(0)}\right) =\nonumber\\
&\left[\begin{array}{rrrrrrrr}
         a& b& b& \frac{x}{y}& \frac{x}{y}& -\frac{x^2}{yz}& \frac{x^2}{byz}& \frac{x^2}{byz}\\
         b& x& y& \frac{x}{y}& z& -x& -1& \frac{x}{z}\\
         b& y& x& z& \frac{x}{y}& -x& \frac{x}{z}& -1\\
         \frac{x}{y}& \frac{x}{y}& z& \frac{x^2}{y^2}& -\frac{z}{y}& -\frac{x^2}{yz}& \frac{x}{y}& \frac{x^2}{y^2z}\\
         \frac{x}{y}& z& \frac{x}{y}& -\frac{z}{y}& \frac{x^2}{y^2}& -\frac{x^2}{yz}& \frac{x^2}{y^2z}& \frac{x}{y}\\
         -\frac{x^2}{yz}& -x& -x& -\frac{x^2}{yz}& -\frac{x^2}{yz}& c& -\frac{x^3}{y^2z^2}& -\frac{x^3}{y^2z^2}\\
         \frac{x^2}{byz}& -1& \frac{x}{z}& \frac{x}{y}& \frac{x^2}{y^2z}& -\frac{x^3}{y^2z^2}& \frac{x^3}{y^2z^2}& \frac{x^2}{yz^2}\\
         \frac{x^2}{byz}& \frac{x}{z}& -1& \frac{x^2}{y^2z}& \frac{x}{y}& -\frac{x^3}{y^2z^2}& \frac{x^2}{yz^2}& \frac{x^3}{y^2z^2}\\
\end{array}\right],
\end{align}
where
\begin{align}
    a &= 2(y + z) + x(2 + x/y)/z + 2x^2/((y + z)(x + yz)) - 1,\\
    b &= -x/y - y - x/z - z,\\
    c &= 2x + x^2(2x/y/z + 3)/y/z - 1.
\end{align}
Constant values of the triplet $(x,y,z)$ can be obtained as a solution of the system of
nonlinear equations:
\begin{equation}
\left\{\begin{aligned}
	x - 3yz/x + y^2z^2(1 - 2x)/x^3 - 2=0,\\
	1/y + x/y/z^2 + 1/z + x/y^2/z - x/(y + z)/(x + yz)=0,\\
	3 + x(1 - y + z)/y/z + (y + (y - 1)z)/x=0.
 \end{aligned}\right.
\end{equation}

Similarly, the second symmetric matrix $Y_{9B}^{(0)}$ reads
\begin{align}
&{\rm core}\left(Y_{9B}^{(0)}\right) =\nonumber\\
&\left[\begin{array}{rrrrrrrr}
           \frac{c^2}{b}  &   a                 &  b               &  c^2             &  \frac{c^2}{d}    &  c               &   \frac{c^2}{a}      &   d         \\
           a              &   \frac{a^2d^2}{c^4}&  \frac{abd}{c^3} &  \frac{ad}{bc}   &  \frac{ad}{bc^2}  &  \frac{abd}{c^2} &   \frac{ad^2}{c^4}   &   \frac{ad}{c^2}    \\
           b              &   \frac{abd}{c^3}   &  \frac{cb^2}{d}  &  \frac{bc^3}{ad} &  \frac{bc}{d}     &  b^2             &   \frac{bd}{c}       &   \frac{d}{c}       \\
           c^2            &   \frac{ad}{bc}     &  \frac{bc^3}{ad} &  \frac{c^3}{d}   &  c                &  \frac{bc^2}{d}  &   \frac{d}{c}        &   \frac{d}{b}       \\
           \frac{c^2}{d}  &   \frac{ad}{bc^2}   &  \frac{bc}{d}    &  c               &  \frac{1}{c}      &  \frac{bc^2}{ad} &   \frac{d}{bc}       &   \frac{d}{c^2}     \\
           c              &   \frac{abd}{c^2}   &  b^2             &  \frac{bc^2}{d}  &  \frac{bc^2}{ad}  &  \frac{b^2}{c}   &   b                  &   \frac{bd}{c^2}    \\
           \frac{c^2}{a}  &   \frac{ad^2}{c^4}  &  \frac{bd}{c}    &  \frac{d}{c}     &  \frac{d}{bc}     &  b               &   \frac{d^2}{bc^2}   &   \frac{d^2}{c^2}   \\
           d              &   \frac{ad}{c^2}    &  \frac{d}{c}     &  \frac{d}{b}     &  \frac{d}{c^2}    &  \frac{bd}{c^2}  &   \frac{d^2}{c^2}    &   \frac{d}{a}       \\
\end{array}\right],
\end{align}
where constant values of $(a,b,c,d)$ can be obtained as a solution of the system of nonlinear equations:
\begin{equation}
\left\{\begin{aligned}
    1 + a + b + c + c^2 + c^2/a + c^2/b + c^2/d + d=0,\\
    (c + d)/c + b^2(c + d)/d +  b(1 + c/d + c^3/a/d + ad/c^3 + d/c)=0,\\
    1 + 1/a + bd/c^3 + d(1 + 1/b + b)/c^2 + d/b/c + (1 + a)d^2/c^4=0,\\
    1 + d^2/c^4 + d/c + b/c + d/c^3 + ad^2/b/c^4 + b/a + d^2/b/c^3 + d/c^2=0.
\end{aligned}\right.
\end{equation}

Besides, the Sinkhorn procedure can precisely generate isolated matrices $Y_9$ with $\#\Lambda(Y_9)\in\{201,625\}$ and many other with much bigger
cardinalities of the $\Lambda$-set. Currently, they are analytically intractable.

\begin{table}[ht!]
\centering
\begin{tabular}{||l|c|l|l|l|c||}
 matrix &  ${\bf d}$ &  $\#\Lambda $ & source & comment & status\\
\hline
$B_{9}^{(0)}$          & 0 & 6   & \texttt{S}  & ${\rm B}\mathbb{H}(9,6)$ & known\\
\hline
$Y_{9A}^{(0)}$         & 0 & 76  & \texttt{S}  & $Y_{9A}^{(0)}=Y_{9A}^{(0){\rm T}}$ & new\\
$Y_{9B}^{(0)}$         & 0 & 89  & \texttt{SV} & $Y_{9B}^{(0)}=Y_{9B}^{(0){\rm T}}$ & new\\
$Y_{9C}^{(0)}$         & 0 & 105 & \texttt{S}   & $Y_{9C}^{(0)}=Y_{9C}^{(0){\rm T}}$ cf.~\eqref{Y9} & new\\
\hline
... & ... & ... & ... & ... & ...\\
\end{tabular}
\caption{Selected outputs for $N=9$ obtained from the Sinkhorn algorithm (\texttt{S}) and $V_N$ family (\texttt{V}) defined in Section~\ref{CHM_VN}.}
\label{tab:CHM_9}
\end{table}


\subsection{\texorpdfstring{Set $\mathbb{H}(10)$}{}}

Complex Hadamard matrices of order $N=10$ reveal even more diversity of new examples.

\begin{table}[ht!]
\centering
\begin{tabular}{||l|c|l|l|l|c||}
 matrix &  ${\bf d}$ &  $\#\Lambda $ & source & comment & status\\
\hline
$B_{10}^{(0)}$                    & 0 & 6   & \texttt{S}  & ${\rm B}\mathbb{H}(10,6)$ & known\\
$N_{10A}^{(0)}$                   & 0 & 9   & \texttt{S}  & - & known\\
\hline
$V_{10A}^{(0)}$                   & 0 & 109 & \texttt{V}  & - & new\\
$V_{10B}^{(0)}$                   & 0 & 134 & \texttt{V}  & - & new\\
$V_{10C}^{(0)}$                   & 0 & 251 & \texttt{V}  & - & new\\
\hline
$Y_{10A}^{(0)}$                   & 0 & 99  & \texttt{S}  & $Y_{10A}^{(0)}=Y_{10A}^{(0){\rm T}}$ & new\\
$Y_{10B}^{(0)}$                   & 0 & 143 & \texttt{S}  & - & new\\
$Y_{10C}^{(1)}(p_1)$              & 1 & 472 & \texttt{S}  & nonaffine family & new\\
$Y_{10D}^{(2)}(p_1,p_2)$          & 2 & 76  & \texttt{S}  & affine family & new\\
$Y_{10E}^{(3)}(p_1,p_2,p_3)$      & 3 & 278 & \texttt{S}  & nonaffine family & new\\
$Y_{10F}^{(4)}(p_1,p_2,p_3,p_4)$  & 4 & 114 & \texttt{S}  & nonaffine family & new\\
\hline
... & ... & ... & ... & ... & ...\\
\end{tabular}
\caption{Selected outputs for $N=10$ obtained from the Sinkhorn algorithm (\texttt{S}) and $V_N$ family (\texttt{V}) defined in Section~\ref{CHM_VN}.
As in the previous cases there are new representatives from $\mathbb{H}(10)$ and the table is vastly incomplete.}
\label{tab:CHM_10}
\end{table}

Symmetric and isolated matrix $Y_{10A}^{(0)}$ reads
\begin{equation}
Y_{10A}^{(0)} = \left[\begin{array}{rrrrrrrrrr}
	1  &  1   &  1    &  1    &  1   &  1           &  1                            &  1   &  1        &  1     \\
	1  &  f                    &  \frac{1}{d}                &  \frac{1}{d}              &  a                        &  a                      &  ba                           &  c                      &  c                        &  1     \\
	1  &  \frac{1}{d}          &  \frac{1}{ad^2}             &  \frac{1}{d^2}            &  \frac{1}{d}              &  a                      &  \frac{e}{c^2d^2}             &  \frac{c^2}{e}          &  \frac{c}{d}              &  \frac{1}{cd} \\
	1  &  \frac{1}{d}          &  \frac{1}{d^2}              &  \frac{1}{cd^2}           &  \frac{a}{cd}             &  \frac{ac}{e}           &  \frac{b}{d}                  &  c                      &  \frac{bc^2}{e}           &  \frac{1}{d}   \\
	1  &  a                    &  \frac{1}{d}                &  \frac{a}{cd}             &  \frac{ae}{bc}            &  a^2                    &  a^2b                         &  a                      &  \frac{e}{b}              &  \frac{1}{b}   \\
	1  &  a                    &  a                          &  \frac{ac}{e}             &  a^2                      &  \frac{a^2}{c}          &  \frac{ae}{c}                 &  ac                     &  c                        &  \frac{a}{c}  \\
	1  &  ba                   &  \frac{e}{c^2d^2}           &  \frac{b}{d}              &  a^2b                     &  \frac{ae}{c}           &  \frac{e^2}{c^2d^2}           &  e                      &  ba                       &  b     \\
	1  &  c                    &  \frac{c^2}{e}              &  c                        &  a                        &  ac                     &  e                            &  \frac{c^2}{a}          &  c^2                      &  \frac{c}{a}  \\
	1  &  c                    &  \frac{c}{d}                &  \frac{bc^2}{e}           &  \frac{e}{b}              &  c                      &  ba                           &  c^2                    &  \frac{c^3d}{e}           &  cd   \\
	1  &  1                    &  \frac{1}{cd}               &  \frac{1}{d}              &  \frac{1}{b}              &  \frac{a}{c}            &  b                            &  \frac{c}{a}            &  cd                       &  d    
\end{array}\right],
\end{equation}
with
    $e = abcd$ and
    $f = -a(2 + b) - 2(1 + d + cd)/d$,
where constant values of $(a,b,c,d)$ can be obtained as a solution of the system of nonlinear equations:
\begin{equation}
\left\{\begin{aligned}
    1 + a + cd/b + ad(2 + 1/c + c + d) + a^2d(b + cd)/c=0,\\
    2 + 1/b + b + a/c + c/a + 1/d + 1/c/d + d + cd=0,\\
    a^2bcd(1 + b + d) + c(b + d + bd) + ab(1 + cd)^2=0,\\
    c + bd((1 + a)(1 + c)(a + c) + a^2bcd)=0.
\end{aligned}\right.
\end{equation}

In turn, matrix $Y_{10B}^{(0)}$ is slightly more complicated. One must solve a system with at most eight unknowns to obtain an isolated point with exactly $143$ Haagerup invariants. Interestingly, the solution space contains also matrices whose defects and $\#\Lambda$'s exhibit a close resemblance~\cite{CHM_catalogue} to $D_{10}$, $N_{10A}$ and $N_{10B}^{(3)}$.

Finally, we present only one representative of four families $Y_{10C,D,E,F}^{(\delta)}$,
namely the affine one. Due to purely practical reasons, it is more convenient that the remaining complex examples are only published in the form of
Matlab and Mathematica scripts on Github~\cite{GITHUB} and in the Catalog of CHM~\cite{CHM_catalogue}.
Hence, matrix $Y_{10C}^{(2)}(p_1,p_2)$ reads
\begin{align}
&{\rm core}\left(Y_{10C}^{(2)}(p_1,p_2)\right) =\nonumber\\
&\left[\begin{array}{rrrrrrrrr}
          1 & -i   &  i   &   i    & -i   &   \omega^4   &  \omega^4   &  \omega^8   &  \omega^8\\
          1 &  i   & -1   &  -1    &  i   &   \omega^{11}  &  \omega^{11}  &  \omega^7   &  \omega^7\\
          i &  \omega^8 &  \omega^{10}&   \omega^7  &  \omega^{11}&   a     & -a     &  \omega^4   &  \omega^4\\
          i &  \omega^4 &  \omega^2 &   \omega^{11} &  \omega^7 &   \omega^8   &  \omega^8   &  b     & -b\\
         -i &  \omega^8 &  \omega^4 &   \omega^7  &  \omega^5 &   ai   & -ai   &  \omega     &  \omega\\
         -i &  \omega^4 &  \omega^8 &   \omega^{11} &  \omega   &   \omega^5   &  \omega^5   &  bi   & -bi\\
         -1 &  1   & -1   &   1    & -1   &   a\omega^{11}& -a\omega^{11}&  b\omega^7 & -b\omega^7\\
         -1 & -i   & -i   &   i    &  i   &   a\omega^7 & -a\omega^7 &  b\omega^{11}& -b\omega^{11}\\
         -1 &  i   &  1   &  -1    & -i   &  -a     &  a     & -b     &  b\\
\end{array}\right],
\end{align}
where $\omega=\exp\{i\pi/6\}$ and
$a = \exp\{2i\pi p_1\}$, $b = \exp\{2i\pi p_2\}$
are two independent parameters with phases $p_j\in[0,1)$.


\subsection{\texorpdfstring{Set $\mathbb{H}(11)$}{}}

Complex Hadamard matrices of order $N=11$ are exceptional.
So far there is no known any single example of a nonisolated CHM in $\mathbb{H}(11)$ and no reasonable explanation has been proposed that such a thing should not exist.
Here we confirm this interesting character of $\mathbb{H}(11)$ presenting several potentially new isolated matrices.

Additionally, every known $H\in\mathbb{H}(11)$, that is: $F_{11}^{(0)}$, $C_{11\Sigma}^{(0)}$, $N_{11\Sigma}^{(0)}$ and~\cite{SzPhD} $Q_{11\Sigma}^{(0)}$ can be brought to the symmetric form $H=H^{\rm T}$. Generic symmetric matrix of dimension $N$ in its dephased form depends on $\tau(N-1)$ unimodular parameters, where $\tau(N)=\binom{N}{2}$ is a triangular number. Maximal number of different Haagerup invariants for such a matrix reads
$\max\{\#\Lambda\}=1+\tau(N)+\tau^2(N)$. For $N=11$ it is $3081$, and this number also appears among the others in the set of discovered matrices.

Similarly as in the case of $N=7$, random seeds supplied to the \texttt{sinkhorn} procedure
recover the Fourier matrix $F_{11}^{(0)}$, $Q_{11}^{(0)}$, a Butson matrix ${\rm B}\mathbb{H}(11,22)$ and probably other aforementioned $11$-dimensional CHM.
Apart from these matrices, we can also observe two inequivalent isolated
solutions from the family $V_N$~\eqref{UC_LS_undephased} characterized
by $\#\Lambda(V_{11\Sigma})\in\{161,331\}$, both persymmetric (it is enough to change the order of columns to restore symmetry). Moreover, the Sinkhorn algorithm finds plenty of other examples with
\begin{equation}
\#\Lambda(Y_{11\Sigma})\in\{191, 323, 425, 751, 975, 1457, 1561, ..., 3081\},
\end{equation}
all of them being symmetric too.
Both families are symbolically identified by $\Sigma\in\{A,B,C,...\}$.
Peculiar values of the Haagerup sets tentatively qualify each $V_{11\Sigma}^{(0)}$ and $Y_{11\Sigma}^{(0)}$
as new isolated and non-Butson candidates for the elements of the set $\mathbb{H}(11)$.
As another open problem, we leave all of them for a future analytic investigation, along with two questions to be resolved; does there exist any family in $\mathbb{H}(11)$, or is it possible to find a new class which cannot be symmetrized?

There is one more example that can be described analytically. This is the matrix $L_{11}^{(0)}$, a member of the family~\eqref{LS_pattern}, which also can be brought to the symmetric form by reshuffling its columns. With some effort,
one can strictly solve the unitarity constraints for this matrix; Eq.~\eqref{TRIGONOMETRIC_LN}, and write
\begin{proposition}
Matrix $L_{11}^{(0)}$ being a solution of the system of equations~\eqref{TRIGONOMETRIC_LN} is an isolated example of a non-Butson CHM for $N=11$.
\end{proposition}
At this stage we do not know how to simplify the formulas describing $L_{11}^{(0)}$. 
They are too complicated to be listed in any legible form. Again, we refer the Reader
to~\cite{GITHUB}.


\subsection{\texorpdfstring{Set $\mathbb{H}(N)$ for $N>11$}{}}

Analytical approach beyond $N=11$ seems to be out of reach. Already for $N=9$, $10$ and $11$
the formulas are ridiculously overcomplex and classification of matrices of this order
requires entirely new methods.
Let us only mention about few possible candidates of orders up to $N=16$.

Family $V_{12}$~\eqref{UC_LS_undephased} contains at least four inequivalent and isolated solutions:
$\#\Lambda(V_{12A}^{(0)})=58$,
$\#\Lambda(V_{12B}^{(0)})=78$, $\#\Lambda(V_{12C}^{(0)})=189$ and $\#\Lambda(V_{12D}^{(0)})=230$.

From $V_{13}$ one obtains
$\#\Lambda(V_{13A}^{(0)})=49$, $\#\Lambda(V_{13B}^{(0)})=95$, $\#\Lambda(V_{13C}^{(0)})=265$
and $\#\Lambda(V_{13D}^{(0)})=547$. Additionally, postulating the matrix $Y_{13}$ to have the
following block core with circulant blocks
\begin{equation}
{\rm core}(Y_{13})=\left[
\begin{array}{lll|lll}
A & B & C & D & E & F\\
B & C & A & E & F & D\\
C & A & B & F & D & E\\
\hline
D & E & F & C^* & A^* & B^*\\
E & F & D & A^* & B^* & C^*\\
F & D & E & B^* & C^* & A^*
\end{array}\right],
\end{equation}
where, each letter $\{A,B,C,D,E,F\}$ corresponds to a structure defined in~\eqref{2x2_X},
one can write unitarity constraints that can be reduced to only six nonlinear equations:
\begin{equation}
\left\{
\begin{array}{r}
a^2+b^2+c^2+d^2+e^2+f^2+c.c.=-1,\\
a/c+b/a+c/b+d/f+e/d+f/e+c.c.=-1,\\
ac+ab+bc+df+de+ef+c.c.=-1,\\
ae+bf+cd+d/a+e/b+f/c+c.c.=-1,\\
ad+be+cf+a/e+b/f+c/d+c.c.=-1,\\
af+bd+ce+d/b+e/c+f/a+c.c.=-1,
\end{array}\label{UC_X13}
\right.
\end{equation}
with $c.c.$ denoting complex conjugate of the preceding terms.
This gives rise to at least two matrices;
one is equivalent to the symmetric variant of Fourier $F_{13}^{(0)}$, while another isolated case will be
called $Y_{13}^{(0)}$, such that $Y_{13}^{(0)}\not\in{\rm B}\mathbb{H}(13,q)$ for any $1<q\leqslant 2^{16}$, and
$\#\Lambda(Y_{13}^{(0)})=301$. This, in comparison with other (isolated) matrices~\cite{La16, Ha96}:
$\#\Lambda(M_{13A}^{(0)})=6$ and 
$\#\Lambda(C_{13A}^{(0)})=\#\Lambda(C_{13B}^{(0)})=9$,
classifies $Y_{13}^{(0)}$ as a new isolated element of $\mathbb{H}(13)$. Exactly the same matrix can be found in the set $\{\texttt{sinkhorn}(X)\}$ for a random $X\in\mathbb{C}^{13\times 13}$.

Table~\ref{tab:CHM_11_16} briefly summarizes all new examples from the last two sections including new cases for $N\in\{14,15,16\}$.

\begin{table}[ht!]
\centering
\begin{tabular}{||l|l|l||l|l|l||l|l|l||}
 matrix  &  $\#\Lambda $ & source & matrix  &  $\#\Lambda $ & source & matrix  &  $\#\Lambda $ & source\\
 \hline
$L_{11A}^{(0)}$    & 191  & \texttt{SL} &  $V_{13A}^{(0)}$    & 49   & \texttt{V}  & $L_{15}^{(0)}$     & 463  & \texttt{L}  \\
$Y_{11B}^{(0)}$    & 425  & \texttt{S}  &  $V_{13B}^{(0)}$    & 95   & \texttt{V}  & $V_{15A}^{(0)}$    & 343  & \texttt{V}  \\
$Y_{11C}^{(0)}$    & 975  & \texttt{S}  &  $V_{13C}^{(0)}$    & 265  & \texttt{V}  & $V_{15B}^{(0)}$    & 407  & \texttt{V}  \\
$V_{11A}^{(0)}$    & 161  & \texttt{V}  &  $V_{13D}^{(0)}$    & 547  & \texttt{V}  & $V_{15C}^{(0)}$    & 841  & \texttt{V}  \\
$V_{11B}^{(0)}$    & 331  & \texttt{V}  &  $Y_{13}^{(0)}$     & 301  & \texttt{S}  & ...                & ...  & ...         \\
\hline
$V_{12A}^{(0)}$    & 58   & \texttt{V}  &  $V_{14A}^{(0)}$    & 297  & \texttt{V}  & $V_{16A}^{(0)}$    & 449  & \texttt{V}  \\
$V_{12B}^{(0)}$    & 78   & \texttt{V}  &  $V_{14B}^{(0)}$    & 330  & \texttt{V}  & $V_{16B}^{(0)}$    & 538  & \texttt{V}  \\
$V_{12C}^{(0)}$    & 189  & \texttt{V}  &  $V_{14C}^{(0)}$    & 362  & \texttt{V}  & $V_{16C}^{(0)}$    & 1025 & \texttt{V}  \\
$V_{12D}^{(0)}$    & 230  & \texttt{V}  &  $V_{14D}^{(0)}$    & 687  & \texttt{V}  & ...                & ...  & ...         \\
\hline
... & ... & ... & ... & ... & ... & ... & ... & ...
\end{tabular}
\caption{Exemplary outputs for $11\leqslant N\leqslant 16$ obtained from the Sinkhorn algorithm (\texttt{S}),
and families $V_N$ (\texttt{V}) and $L_N$ (\texttt{L}).
This is a selection of only isolated objects.
Family $V_N$ provides also matrices with non-zero defect. All matrices are new, however most of them cannot be presented in any compact form, yet.}
\label{tab:CHM_11_16}
\end{table}

\end{document}